\documentstyle[12pt,axodraw]{article}

\catcode`@=11 \@addtoreset{equation}{section} \catcode`@=12

\begin{document}
\begin{titlepage}
\begin{flushright}\vbox{\begin{tabular}{c}
           TIFR/TH/97-03\\
           January, 1997\\
           hep-ph/yymmdd\\
           \end{tabular}}
\end{flushright}
\begin{center}
   {\large \bf  Thermal Effects on\\
                Two-photon Decays of Pseudo-scalars}
\end{center}
\bigskip
\begin{center}
   {Sourendu Gupta\footnote{E-mail: sgupta@theory.tifr.res.in}
    and S.\ N.\ Nayak\footnote{E-mail: nayak@theory.tifr.res.in}\\
    Theory Group, Tata Institute of Fundamental Research,\\
    Homi Bhabha Road, Bombay 400005, India.}
\end{center}
\bigskip
\begin{abstract}
We study the effect of finite temperatures and Fermion density
on the effective pseudo-scalar-photon vertex induced by the
triangle diagram. The manifestly covariant calculations show
that when the pseudo-scalar mass is much less than the temperature,
then there is a large enhancement of the decay rate. Alternatively,
when the temperature is much higher than the Fermion mass, or the
Fermion chemical potential is large, the lifetime is enhanced.
Other related processes, and applications of these results to cosmology
and astrophysics, are discussed.
\end{abstract}
\end{titlepage}

\section{Introduction\label{intro}}

Goldstone Bosons arise in field theories when some global
symmetry is spontaneously broken. A solution of the strong CP problem
involves the breaking of Peccei-Quinn symmetry and gives rise to a
pseudo-scalar called the axion ($A$) \cite{axion}, which becomes
massive through mixing with the pion. Axions have couplings to charged
Fermions and are candidate dark matter which could play an important
role in cosmology \cite{cosmo}. The physics of axions involves two
intrinsic mass scales--- the axion mass, $M$, and the Fermion mass,
$m$. If axions participate in physics inside a heat bath with temperature,
$T$, larger than any one of (or both) these scales, then we may expect
interesting and non-trivial thermal effects. Since the mass of the axion
can be much less than the cosmic microwave background temperature ($T_b=
2.7$ K), even the coldest environment in the universe may be thermal in
this sense. The purpose of this paper is to identify the main sources of
non-trivial thermal effects.

We compute the decay rate of the axion into two photons through the
triangle diagram and find two sources of
thermal effects. Stimulated emission of final state photons enhances the
decay width of the pseudo-scalar, and plays the dominant role in the
physics of axions for $M<T_b$. Pauli blocking of on-shell parts of the
Fermions in the loop reduces the decay width when $T>m$, and may have
interesting consequences in the physics of supernov\ae. These changes
in the effective $A\gamma\gamma$ vertices also contribute to the Primakoff
effect, $\gamma e\to Ae$ through a $t$-channel photon exchange. Thermal
effects can also boost the multi-photon channels to a very large extent
and may require a resummation of this whole class of processes.

One bit of physics is worth emphasizing. At $T=0$ the decay width, $\Gamma$,
depends on all the scalars in the problem. These are the masses $M$ and $m$.
The momenta of the decay photons, $k_1$ and $k_2$ ($k_1+k_2=q$, the momentum
of $A$) do not contribute anything else, since $k_1\cdot k_2=M^2$. There are
major changes in the kinematics at finite temperature, since the heat bath
selects out a preferred frame. A covariant description can be retained in
finite temperature field theory, provided an extra vector is introduced into
the problem--- the velocity of the heat-bath with respect to the frame in
which one chooses to work, $u$ ($u^2=1$) \cite{weldon}. The amplitude may
now depend on all the scalars in the problem. There are four new scalars---
the temperature $T$, the Fermion chemical potential $\mu$, and two
scalar-products, which can be chosen as $q\cdot u$ and $k_1\cdot u$. Since
$\Gamma$ is obtained after integration over the two-photon phase
space, the dependence on $k_1\cdot u$ drops out. Consequently, $\Gamma$ is
a function of $M$, $m$, $T$, $\mu$ and $q\cdot u$. This last quantity is
the energy of $A$ in the rest frame of the heat-bath.

A simple representation of the new invariants facilitates the discussions
in the later sections. Evaluate them in the rest frame of $A$, $q=(M,0,0,0)$.
Align $u$ along the $z$-direction and write $u=(\gamma,0,0,\beta)$. Since
the energy of each photon is $M/2$, and the momenta balance,
\begin{equation}\begin{array}{rl}
   q\cdot u\;=&\;M\gamma, \\
   k_1\cdot u\;=&\;{1\over2}M\gamma(1-\beta\cos\chi), \\
   k_2\cdot u\;=&\;{1\over2}M\gamma(1+\beta\cos\chi).
\end{array}\label{intro.inv}\end{equation}
It is now obvious that $\gamma$ and $\chi$ are Lorentz invariants. Also,
$\beta=\sqrt{(1-1/\gamma^2)}$ can be interpreted as the boost between the
rest frames of $A$ and the heat-bath.

The plan of this paper is the following. In Section \ref{temp} we present
the calculation of the triangle diagram at $T>0$, compute the decay width,
and discuss some applications of these results. The next section contains
a discussion of other related processes, including the Bose enhancement of
multi-photon decays. Some field-theoretic points are discussed
in the appendix.

\section{The Triangle Diagram at $T>0$.\label{temp}}

In this section we compute the triangle diagram at $T>0$. A limiting
case, $\gamma=1$, has been considered before \cite{prev}. In all physical
applications it is appropriate to consider the limit $M\ll m$. The
photon-Fermion coupling is taken to be $ie\hat\varepsilon(k)$ (for any vector
$p$ we denote $\hat p=\gamma_\mu p^\mu$), where $\varepsilon(k)$ denotes the
polarisation vector of a photon of momentum $k$. The Fermion-$A$ coupling
is $ig\gamma_5$.

We summarise the computation of the decay width at $T=0$. The matrix element,
${\cal M}$, coming from the triangle diagram, has the form 
\begin{equation}
   {\cal M}\;=\; -\,{\displaystyle ge^2\over\displaystyle4\pi^4}
       \varepsilon^\mu(k_1)\varepsilon^\nu(k_2)\,T_{\mu\nu}
                                  f(M,m)\quad{\rm where}\quad
   T_{\mu\nu}=m\epsilon_{\mu\nu\sigma\rho}k_1^\sigma k_2^\rho.
\label{temp.zero}\end{equation}
The Lorentz-scalar, $f$, is a form-factor for the effective $A\gamma\gamma$
vertex and is obtained as an integral over the Fermion-loop momentum. It
is a function only of $M$ and $m$. It is possible to expand $f$ as
\begin{equation}
   f(M,m)\;=\;f_0(m)+z^2 f_1(m)+\cdots,
   \qquad(z={\displaystyle M\over\displaystyle2m}).
\label{temp.form}\end{equation}
A text-book calculation gives $f_0=-\pi^2/m^2$ and $f_1=-\pi^2/3m^2$
\cite{textbk}. The decay width, $\Gamma$, is computed by integrating the
squared matrix element over the Lorentz-invariant phase space of the final
state--- 
\begin{equation}
   \Gamma(M,m)\;=\;\left({\displaystyle{\alpha^2g^2\over16\pi^3}}\right)
       z^2 M.
\label{temp.width}\end{equation}

In the usual formulation of finite-temperature field theory \cite{ftft},
the physical fields (called type 1) are doubled by the addition
of so-called ``thermal ghosts'' (type 2 fields). Each vertex involves
only one type of fields. External legs connect only to type 1 vertices.
Each propagator may connect any two types of vertices and hence becomes a
$2\times2$ matrix which can easily be written down in terms of advanced
and retarded Green's functions along with the Bose or Fermi distributions.
In the triangle diagram all vertices connect to
external legs, and we need to consider only the type 11 Fermion propagators
\cite{dj}
\begin{equation}\begin{array}{rl}
   {\cal D}(p,u) \;=\; (\hat p+m)&
          \biggl[{\displaystyle 1\over\displaystyle p^2-m^2+i\epsilon}
               + 2\pi i\delta(p^2-m^2)\times\\
       & \qquad\left\{\Theta(p_0)F^+(p\cdot u)
      +\Theta(-p_0)F^-(p\cdot u)\right\}\biggr],
\end{array}\label{temp.prop11}\end{equation}
where a regulator has been placed in the denominator of the propagator,
$F^\pm (x)=1/(\exp(|x|\mp\mu)/T+1)$, and $\mu$ is the Fermion chemical
potential. Note that this is the sum of a $T=0$ and a $T>0$ part. It is
convenient to use a diagrammatic notation where the $T=0$ part is denoted by
a line and the $T>0$ part by a line with a cut. Thermal effects on external
legs can be subsumed into Bose and Fermi distributions multiplying the phase
space volume element. We return to this point in the appendix.

\subsection{The Form Factor}
Note that the Dirac structure of the propagator is the same as at $T=0$.
As a result, the tensor structure of the matrix element remains as in
eq.\ (\ref{temp.zero}) and the only thermal effects come from the form
factor $f$ and the photon phase space. The modified form factor is
\begin{equation}
   f\;=\;\int d^4p {\cal D}(p){\cal D}(p-k_1){\cal D}(p-q).
\label{temp.matel}\end{equation}
Separating the $T=0$ and $T>0$ parts of the propagator, the thermal
form factor has eight terms. Apart from the $T=0$ contribution, other terms
can be grouped as 
(a) one term with the thermal part of all three propagators
(b) three terms with the thermal part of any two and
(c) three terms with the thermal part of only one propagator.
For $M<2m$ and $m>0$, the three mass-shell conditions in (a)
leave no phase-space volume to the integral and hence this contribution is
identically zero. For the same reason each term in (b) is also zero.
Consequently, the thermal contribution comes only from (c), written in
a diagrammatic notation as
\begin{equation}\begin{array}{rl}
   f\;=&\; 2\left(
      \begin{picture}(70,30)(-10,-2)
      \SetPFont{}{8}
      \Line(0,0)(34,20) \Text(16,14)[rb]{$p$}
      \Line(34,20)(34,-20) \Text(35,21)[lb]{$\mu,k_1$}
      \Line(34,-20)(0,0) \Text(35,-21)[lt]{$\nu,k_2$}
      \Line(31,0)(37,0) \Text(-1,0)[r]{$q$}
      \LongArrowArcn(22.66,0)(8,60,180)
      \end{picture}    +
      \begin{picture}(70,30)(-10,-2)
      \SetPFont{}{8}
      \Line(0,0)(34,20) \Text(16,14)[rb]{$p$}
      \Line(34,20)(34,-20) \Text(35,21)[lb]{$\mu,k_1$}
      \Line(34,-20)(0,0) \Text(35,-21)[lt]{$\nu,k_2$}
      \Line(15.03,13.48)(18.97,6.52) \Text(-1,0)[r]{$q$}
      \LongArrowArcn(22.66,0)(8,60,180)
      \end{picture}    +
      \begin{picture}(70,30)(-10,-2)
      \SetPFont{}{8}
      \Line(0,0)(34,20) \Text(16,14)[rb]{$p$}
      \Line(34,20)(34,-20) \Text(35,21)[lb]{$\mu,k_1$}
      \Line(34,-20)(0,0) \Text(35,-21)[lt]{$\nu,k_2$}
      \Line(15.03,-13.48)(18.97,-6.52) \Text(-1,0)[r]{$q$}
      \LongArrowArcn(22.66,0)(8,60,180)
      \end{picture}
      \right)\\ \\ \;=&\;2(J_1+J_2+J_3).
\end{array}\label{integ.result}\end{equation}
The factor of two comes from the diagrams obtained
by simultaneous interchange of $\mu$, $\nu$ and $k_1$, $k_2$.

Each of the integrals $J_i$ can be reduced to a Lorentz invariant
one-dimensional Fermi integral. We show some of the details for $J_1$,
to demonstrate that the computation can be performed covariantly.
After shifting $p\to p+k_1$, and introducing a resolution of identity
as a sum over positive and negative energy $\Theta$-function, $J_1$
reduces to a simple integral over a real particle phase space
\begin{equation}
   J_1\;=\;-\pi\int{\displaystyle d^3p\over\displaystyle2p_0}(F^-(x)+F^+(x))
     {\displaystyle1\over\displaystyle p\cdot k_1\,p\cdot k_2}
\label{integ.j1}\end{equation}
This integral is finite and well-defined for $m>0$, and, being a
Lorentz-scalar, may be evaluated in any convenient frame. We
use the frame where $u=(1,0,0,0)$. In this frame $p_0=p\cdot u$ is
a Lorentz-scalar. As a result, ${\bf p}^2=p_0^2-m^2$ is also invariant,
and all the manipulations shown here are explicitly Lorentz invariant.

First we use the Feynman trick to write
\begin{equation}
   J_1=-{\displaystyle{\pi\over2}}\int
           {\displaystyle{d{\bf p}\,{\bf p}^2\over p_0}}
              \left[F^-(p_0)+F^+(p_0)\right]\Omega_1,
   \quad{\rm where}\quad
   \Omega_1=\int{\displaystyle{d\alpha d\Omega\over(p\cdot V)^2}}
\label{integ.j1next}\end{equation}
Here $\alpha$ is the Feynman parameter ($0\le\alpha\le1$) and
$V=\alpha k_1+(1-\alpha)k_2$. We choose the orientation of the frame
by setting $V=(V_0,0,0,V_3)$. Since $V_0=V\cdot u$, both $V_0$ and $V_3$ are
Lorentz-invariant. The angular integral is easily performed, leaving
$\Omega_1=(4\pi/M^2)\int d\alpha/Q_1(\alpha)$, where the $Q_1(\alpha)$ is
a quadratic in $\alpha$ with Lorentz-invariant coefficients. It is easy to
check that the two roots of $Q_1$ do not lie in the range of integration.
Then the integral over $\alpha$ is easily performed. With the notation
\begin{equation}
   {\cal P}^2\;=\;{\bf p}^2 + m^2\gamma^2(1-\beta^2\cos^2\chi),
\label{integ.mnem}\end{equation}
we have the compact expression
\begin{equation}
   J_1\;=\;-\,
         {\displaystyle{4\pi^2\over M^2}}\int_0^\infty
             {\displaystyle{d{\bf p}\,{\bf p}\over p_0{\cal P}}}
         \left[F^-(p_0)+F^+(p_0)\right]
      \log\left({\displaystyle{{\cal P}+{\bf p}\over
                      {\cal P}-{\bf p}}}\right).
\label{integ.j1fin}\end{equation}
$J_2$ and $J_3$  can be obtained in a similiar fashion. 

The $T>0$ result may be exhibited in a form similiar to that in
eq.\ (\ref{temp.form}) by expanding the integrals in a series in $z=M/2m$.
Each of the integrals $J_i$ ($i=1,2,3$) has an overall factor of $1/M^2$.
The only $M$ dependence of $J_1$ is in this factor. The expansion of $J_2+
J_3$, starts at order $1/z$. A straightforward check shows that the first
term of the expansion precisely cancels $J_1$. The first non-zero term is
of order $z^0$ and this is followed by a series in $z^2$. All these terms
in the expansion come entirely from $J_2+J_3$.

\begin{figure}
\vskip12truecm
\includegraphics{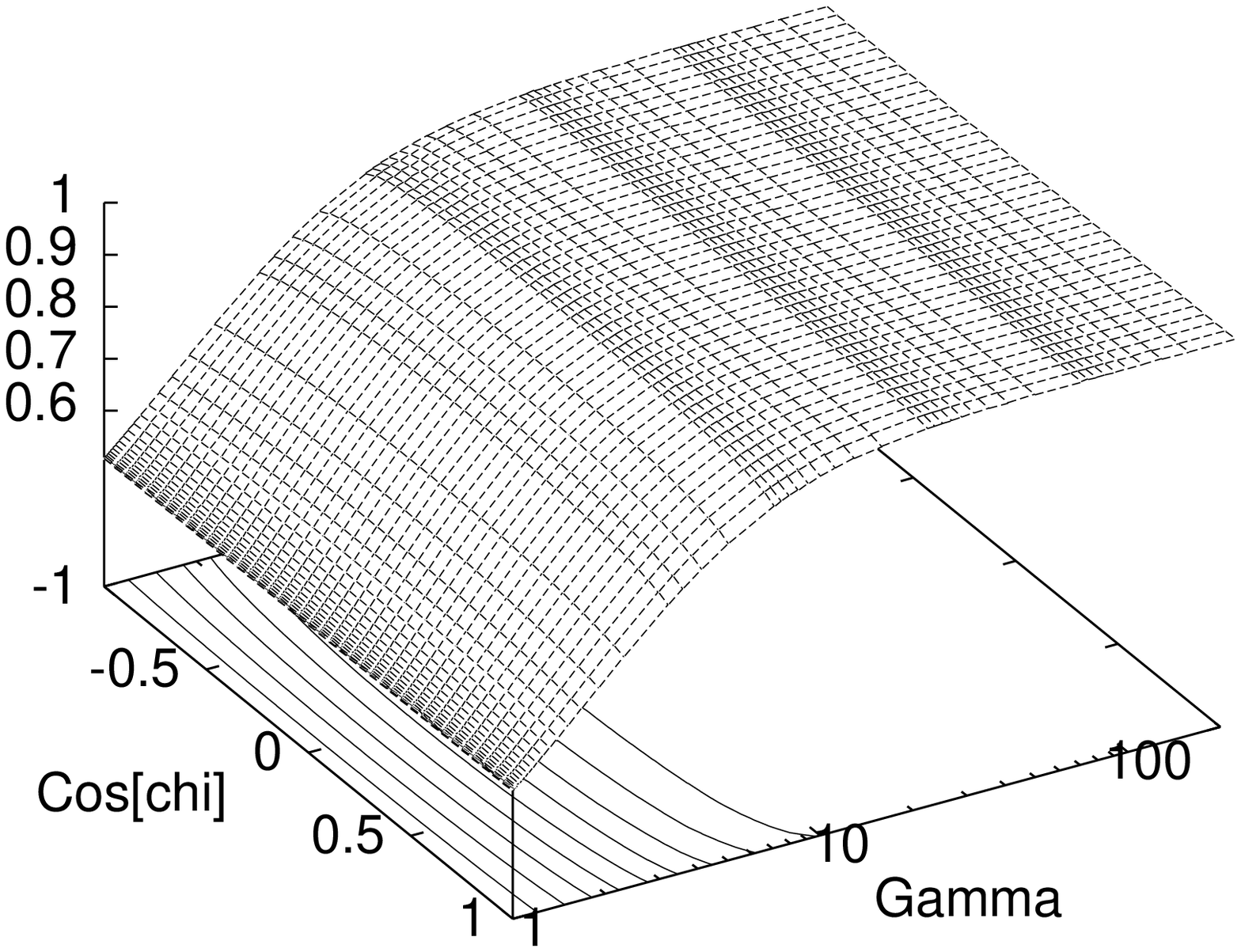}
\includegraphics{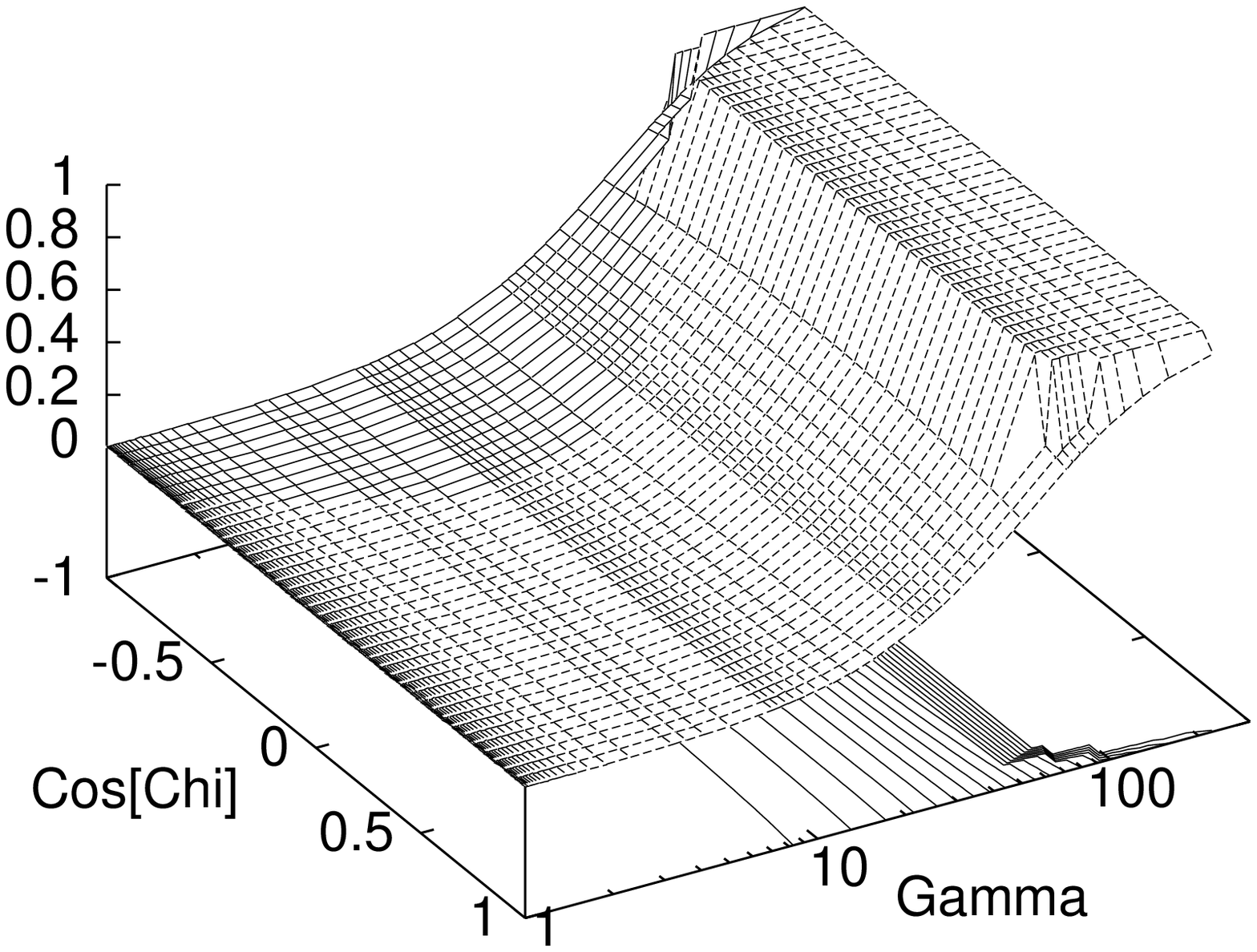}
\caption[dummy]{The form-factor $|f_0m^2/\pi^2|$ shown as a function of
   $\gamma$ and $\cos\chi$ for values of $T$ and $\mu$ appropriate to
   a supernova core ($T=40$ MeV, $\mu=450$ MeV; lower panel) and the
   neutrino-sphere in a supernova ($T=1$ MeV, $\mu=0$; upper panel).
   Isolines drawn on the $\gamma$-$\cos\chi$ plane correspond to
   function values increasing in steps of $0.05$ upto a largest value
   (on the right) of $0.95$.}
\label{fg.temp.form}\end{figure}

We give here the final result for the full $T$-dependent $f_0$---
\begin{equation}\begin{array}{rl}
 &f_0(M,m,\mu,T,\gamma)\;=\;-\,{\displaystyle\pi^2\over\displaystyle m^2}
        +{\displaystyle\pi^2\gamma^2\over\displaystyle m^2}\int_0^\infty
       {\displaystyle dr\,r\over\displaystyle r_0R^5}
       \bigl[F^-(\zeta r_0)+F^+(\zeta r_0)\bigr]\\
     &\qquad\qquad\quad\times
      \biggl[\left(2r_0^2-\gamma^2\beta^2\sin^2\chi\right)
        \biggl\{(1+\bar\beta^2)
            \log\left({\displaystyle R+r\over\displaystyle R-r}\right)\\
     &\qquad\qquad\qquad\qquad\qquad\qquad\qquad\qquad\qquad
             -2\bar\beta\log\left(
           {\displaystyle R+r\bar\beta\over\displaystyle R-r\bar\beta}\right)
            \biggr\}\\
     &\qquad\qquad-
        {\displaystyle2rR\over\displaystyle r^2+\gamma^2}
      \left(2r_0^2\overline\beta^2-(r^2+\gamma^2)\beta^2\sin^2\chi\right)
           \biggr],
\end{array}\label{temp.result}\end{equation}
where $\bar\beta=\beta\cos\chi$, $r_0^2=1+r^2$,
$R^2=r^2+\gamma^2(1-\bar\beta^2)$, and $\zeta=m/T$. This complicated looking
result may be simplified in various limits, or evaluated numerically (see
Figure \ref{fg.temp.form}).

An instructive limit is $\gamma\to1$, {\sl i.\ e.\/}, the special
case when $A$ is at rest in the rest-frame of the heat bath, and $\mu=0$---
\begin{equation}
   f(M,m,T)\;=\;-\,{\displaystyle\pi^2\over\displaystyle m^2}
       +{\displaystyle2\pi^2\over\displaystyle m^2}\int_0^\infty
      {\displaystyle r\,dr\over\displaystyle r_0^4}
     \log\left({\displaystyle r_0+r\over\displaystyle r_0-r}\right)
     {\displaystyle1\over\displaystyle{\rm e}^{\zeta r_0}+1}.
\label{temp.bt0}\end{equation}
This result agrees with previous expressions obtained in this limit \cite{prev}.

It is also interesting to see that in the limit $T\gg m$, the integral
takes on the value $1/2$ and the thermal part of $f$ precisely cancels
the $T=0$ result, giving $f=0$. Thus the decay width vanishes in the
joint limits $\beta\to0$ and $M\ll m\ll T$. The reason is quite simple.
A close perusal of the $T=0$ calculation shows that $f$ picks up a
non-vanishing contribution only from that part of the phase space where
one of the Fermions is on-shell. This contribution is fully Pauli-blocked
in the limit $T\gg m$.

For $m\ll T$ it is useful to perform an expansion of the form factor $f$.
From eq.\ (\ref{temp.bt0}) it is clear that
\begin{equation}
   f\;=\;-\,{\displaystyle\pi^2\zeta\over\displaystyle2m^2}\int_0^\infty
      {\displaystyle r\,dr\over\displaystyle r_0^3}
     \log\left({\displaystyle r_0+r\over
               \displaystyle r_0-r}\right)+{\cal O}(\zeta^2)
    \;\sim\;-0.7707\zeta
       \left({\displaystyle\pi^2\over\displaystyle m^2}\right)
\label{temp.expand}\end{equation}
In the other limit, $m\gg T$, the thermal contribution vanishes exponentially
in $\zeta$, leaving $f$ to take on its $T=0$ value.

The results are even simpler when $A$ travels very fast through the medium,
{\sl i.e.\/}, in the limit $\gamma\to\infty$. In this limit the quadratics
appearing in the definitions of all the $J$'s go as $\gamma^2$. Hence the
thermal contribution vanishes as $1/\gamma^2$, and the matrix element is
equal to its $T=0$ limit.

\subsection{The decay width\label{width}}
Next we construct the decay width for $A\to\gamma\gamma$. In addition
to the matrix element computed in the previous section, we require the
2-body decay phase space for $T>0$. This contains a factor $1+B(k\cdot u)$
for each photon (due to stimulated emission in the heat-bath), where
$B(x)=1/(\exp|x|-1)$ is the Bose distribution function. Then, after the
usual reduction of the 2-body phase space,
\begin{equation}\begin{array}{rl}
  \Gamma(M,m,\mu,T,\gamma)\;=\;{\displaystyle g^2\alpha^2 m^2 M^3\over
             \displaystyle128\pi^7}&
     \int_{-1}^1 d\cos\chi f^2(M,m,\mu,T,\gamma,\chi)\\
        &\qquad\left[1+B(k_1\cdot u)\right] \left[1+B(k_2\cdot u)\right].
\end{array}\label{temp.gamma}\end{equation}
For the value of $f$ calculated in the previous section, the integral has
to be performed numerically for general values of the argument.

When the $\chi$-dependence of the form factor can be neglected, the thermal
effect arises entirely from the phase space factors. In this case the decay
width is
\begin{equation}
  \Gamma(M,m,\mu,T,\gamma)\;=\;
     \displaystyle{g^2\alpha^2m^2M^3f^2\over64\beta\xi\pi^7}
     \left({1\over1-{\rm e}^{-2\xi}}\right)
        \left[\beta\xi-\log\left(
               {{\rm e}^{(1+\beta)\xi}-1\over{\rm e}^{(1-\beta)\xi}-1}\right)
         \right],
\label{temp.weight}\end{equation}
where $\xi=M\gamma/2T$. For $\beta\to0$ and $M\ll T$, we find that $\Gamma$
is the $T=0$ width times $T^2/M^2$. On the other hand, for $T\ll M$ we recover
the $T=0$ result. For general values of the parameters a numerical integration
of the expression in eq.\ (\ref{temp.gamma}) is required.

The physics of the temperature dependence of the decay width can be summarised
as---
\begin{enumerate}
\item
    For $T\ll M\ll m$ or large $\beta$, thermal effects can be neglected and
    the decay width is the same as at $T=0$.
\item
    For small $\beta$ and $M\ll T\ll m$ the dominant effect is the stimulated
    emission of final state photons, leading to a highly enhanced decay rate.
\item
    For small $\beta$ $M\ll m\ll T$ Pauli blocking of the on-shell part of the
    loop determines the physics and the decay rate is reduced.
\end{enumerate}

\subsection{Some Applications}
The simple computations above can be used to bound $M$. A large effect
comes from the fact that $\Gamma\sim M^3$ at $T=0$, whereas $\Gamma\sim
T^2 M$ for $M\le T$. This thermal enhancement of the decay width has an
interesting consequence when $M\ll T_b=0.235\times10^{-3}$ eV. The $T=0$
result, in conjunction with some cosmological arguments, would predict a
bound on $g<c/M^{3/2}$, whereas the $T>0$ results strengthen this 
to $g<c'/\sqrt M$ ($c$ and $c'$ are computable constants).

If $A$ is to be a candidate for cold dark matter, then $\beta\approx0$ and
the lifetime must be greater than the age of the universe, $t_0$ \cite{cosmo}.
Then
\begin{equation}
   g \;\le\; \left({8\pi\overline m\over\alpha M}\right)
             \left(1-{\rm e}^{-M/2T_b}\right)
             \sqrt{\pi\over Mt_0}
\label{cosmo.res0main}\end{equation}
where the average Fermion mass $\overline m^{-2}\;=\;\sum_f m_f^{-2}$
is dominated by the lightest flavour, the electron. Assuming the present age
of the universe to be $15\times10^9$ years, we find that when $M\gg T_b$,
$c=1.2\times10^{-7}\ {\rm eV}^{3/2}$ and when $M\ll T_b$,
$c'=2.5\times10^{-4}\ {\rm eV}^{1/2}$.

The coupling $g$ also can be constrained by the condition that the energy
density of the decay photons is limited by observation of the micro-wave
background, $\rho_\gamma$. We will assume \cite{axion-soft} that $A$ is
produced non-thermally in the early universe. Then we can use the
decay width in the limit $\beta\to0$. The additional photon energy density
due to the decay of $A$ at present is
\begin{equation}
   \Delta\rho_\gamma(t)\;=\;\overline\rho_{\scriptscriptstyle A}(t)
     \left[1-{\rm e}^{-\Gamma(t-t_i)}\right],
\label{cosmo.phden}\end{equation}
where $t_i$ is the production epoch and $\overline\rho_{\scriptscriptstyle A}$
is the energy density which would have remained with $A$ in the absence of
decays. Now $\Delta\rho_\gamma\le\rho_\gamma$, and
$\overline\rho_{\scriptscriptstyle A}$ must be greater than the present
critical density $\rho_{cr}$ \cite{axion-soft}. Since $t_i\ll t$,
we find
\begin{equation}
   g \;\le\; \left({8\pi\overline m\over\alpha M}\right)
             \left(1-{\rm e}^{-M/2T_b}\right)
             \sqrt{-\pi\log(1-\rho_\gamma/z\rho_{cr})\over Mt_0}
\label{cosmo.res1main}\end{equation}
Taking $\rho_{\scriptscriptstyle A}=z\rho_{cr}$ with $z=0.9$, $\rho_\gamma
=0.260{\rm \ eV}/{\rm cm}^3$, $\rho_{cr}=1.054\times10^{-5}h_0^2$
${\rm GeV}/{\rm cm}^3$ and choosing $h_0=0.85$ (to get the weakest bound),
we find that $c=7.1\times10^{-10}\ {\rm eV}^{3/2}$ and
$c'=1.5\times10^{-6}\ {\rm eV}^{1/2}$.

On the other hand, if we assume that $\Delta\rho_\gamma$ at the present
epoch is smaller than the known error induced through the error on the
COBE measurement of $T_b$, then the constraints are harder by a factor
of $0.04$. This gives $c=2.8\times10^{-11}\ {\rm eV}^{3/2}$ and
$c'=6\times10^{-8}\ {\rm eV}^{1/2}$. In all these cases the bound on
$g$ is sharpened by two orders of magnitude through the leading order
diagram alone. Although this does not place very stringent restrictions
on axion models, in the next section we consider higher order diagrams and
argue that they may lead to even sharper bounds.

Apart from axion models, these computations can also be used to bound
the couplings and masses of any other pseudo-scalar with a $\gamma_5$
vertex to charged Fermions. One such model, which is not ruled out by
data, is a singlet-triplet Majoron model \cite{anjan}. This Majoron
is a Goldstone Boson for Lepton number breaking and may be given a mass
through couplings to gravity \cite{mahapatra}. Since this mechanism
typically produces low masses, the thermal bound is applicable.

Another class of limits arises from consideration of the cooling rates
of stars from emission of pseudo-scalars. Our computation of the form
factor is applicable to this situation, not only through the direct
two-photon channel, but also through the Primakoff process. Thermal
effects in most stars are rather small, due to the fact that $T\ll m$.
However, they can be important in supernov\ae{} since the $A\gamma\gamma$
vertex is suppressed (see Figure \ref{fg.temp.form}). This may affect
bounds on axions from supernova cooling rates\cite{ellis,turner}.

\section{Other Processes.\label{disc}}

\subsection{Multi-photon decays}
We have computed thermal effects on the decay width of a pseudo-scalar
(of mass $M$) in the two-photon channel through the triangle diagram
with charged Fermions of mass $m$ circulating in the loop.
We found a strong increase in the decay width due to
Bose enhancement of the final state photons for $M<T\ll m$, leading
to a sharpening of the bound on the coupling between the pseudo-scalar
and Fermions by two orders of magnitude for $M=10^{-5}$ eV.
Multi-photon decays are likely to yield more stringent bounds.

The total width of the axion into multi-photon channels at
$T=0$ can be written as the perturbation series
\begin{equation}
   \Gamma \;=\; \alpha^2\Gamma_2 + \alpha^4\Gamma_4 + \cdots
\label{disc.pert}\end{equation}
Since $\alpha^2\approx10^{-4}$, the series converges extremely fast and
even the second term can be neglected. However, the situation changes at
finite temperature. The higher order terms, involving multi-photon decays
are enhanced due to Bose factors on each external photon leg, and some of
these may compensate the damping due to the fine structure constant.

Examine the case of a pseudo-scalar decaying into $2n$ photons, where
the $i$-th photon momentum is $k_i$. When each $k_i\sim M/2n$, the Bose
factors go as $(2nT/M)^{2n}$. As a result, the contribution of this
process to the total decay width is
\begin{equation}
   \alpha^{2n}\Gamma_{2n} \;=\; g^2\alpha^{2n}{c_n\over(2n)!}
         \left({2nT\over M}\right)^{2n} M \left({M\over m}\right)^{2n}.
\label{disc.series}\end{equation}
The last factor comes from propagators and the Dirac trace, the
factorial is due to photon counting and $c_n$ is essentially the
dimensionless contribution due to the Fermion loop integral. Clearly,
this region of phase space is not very important at higher orders
since $T\ll m$.

However, in the part of phase space where one or more $k_i\to0$ the
Bose factors can become arbitrarily large. This can happen for $n\ge2$.
It is easy to check that the real $2n$ photon emission diagrams give
non-vanishing amplitudes in this region of phase space. This is not the
usual Bloch-Nordsieck problem, since the $T=0$ width is perfectly well
defined and insensitive to the infra-red (as long as $m>0$). We expect
these putative divergences to cancel when thermal ghosts and virtual
corrections are taken into account. The result is that, the decay width
at $T>0$ is infra-red sensitive and hence must be resummed over all
numbers of photons. Demonstrating the cancellation and extracting the
finite parts, prior to summing the series then requires the full power
of thermal field theory. This work is under progress and will be reported
elsewhere.

\subsection{Stellar Cooling}

Stellar cooling arguments are dependent on the coupling of axions to
matter. For small couplings axions stream out of the star, carrying
energy. At larger couplings, the mean-free path may be smaller than
the radius of the star, and an axio-sphere may form. Axions escape
only from the surface of the axio-sphere. 

We have seen that the axion-photon form factor decreases significantly
at temperatures appropriate to the core of supernov\ae. This does not
have a significant effect on the cooling rate due to low-mass axions,
since the decrease affects only the low-energy end of the spectrum.
For small $M$ the energy loss due to low-energy axions is negligible
and large changes in production rate can be tolerated.

The main effect is seen for $M$ large enough that an axio-sphere may
be formed. Apart from the decrease in the form factor, one should also
take into account changes in the pion decay constant, $f_\pi$, and the
pion mass, $m_\pi$, since they enter the axion coupling. At $T\approx
40$ MeV, chiral perturbation theory shows some decrease in both these
quantities \cite{leut}. Further, suppression of interaction rates
of low-energy axions with leptons or hadrons, due to Pauli blocking,
increases the mean-free path of low-energy axions. The net effect is
that low-energy axions may escape from the core. At the same time,
high-energy axions are converted to low energies by processes such
as $Ae\to Ae$ (Compton). The inverse process of boosting the energy
of the axions is suppressed by Pauli blocking and the falling density
of low-energy axions. This may lead to a destabilisation of the
axio-sphere. Quantitative estimates of the relaxation time for this
instability and the equilibrium phase space distribution of axions
will be detailed elsewhere.

Even in parts of parameter space where this instability is negligible,
estimates of the axio-sphere radius and the critical coupling at which
it forms necessitate a full thermal field theory computation. The radius
of the axio-sphere is closely related to the relaxation time (and the
mean free path length) of the axion. The relaxation rate is the difference
of the production and decay rates (see the appendix) and is given by the
imaginary part of the axion two-point function. This quantity is
strongly affected by hard thermal loops, if the mass of the axion is
less than $\sqrt\alpha T$. Assuming $T\approx40$ MeV, treatments of
axions with mass less than about 4 MeV in the supernova core require
hard thermal loop resummation.

\appendix
\section{Some Field-theoretic Niceties.\label{htl}}
In Section \ref{temp} we have written down rules for computation of
$T>0$ decay widths which look very similiar to those at $T=0$. The
formalism for doing this is developed in \cite{cutting,welcut}. We
summarise the results in this appendix and show that our techniques
are justified.

In general one computes decay rates by writing down cutting rules
for loop contributions to a two-point function. At $T>0$ loop diagrams
require the matrix propagator \cite{ftft}
\begin{equation}\begin{array}{rl}
   iG(p)\;=&\;U(T,p)\left(\matrix{S(p)&0\\0&S^*(p)}\right)U(T,p),\\
   S(p)\;=&\;{\displaystyle i\over\displaystyle p^2-m^2+i\epsilon},\\
   U(T,p)\;=&\;B(p\cdot u)\left(\matrix{1&{1\over2}\exp|p\cdot u|\\
                             {1\over2}\exp|p\cdot u|&1}\right),\\
   iG^\pm(p)\;=&\;2\pi\delta(p^2-m^2)\left[\Theta(\pm p_0)+B(p\cdot u)\right].
\end{array}\label{htl.prop}\end{equation}
These rules for Bosons can be generalised in a similiar form to Fermions
and Gauge Bosons. At $T=0$ the matrix is diagonal and only the
$\Theta(\pm p_0)$ term appears on $G^\pm$.

The $T=0$ notion of cutting a line is generalised at $T>0$ to the notion
of ``circling'' a vertex \cite{cutting}. If a $G_{11}$ line is circled
only at one end, then it is replaced by $G^+$ if the momentum flows to
the circled vertex ($G^-$ otherwise). For a $G_{22}$ line the rule is
reversed, and the off-diagonal lines are left untouched. Since the
off-diagonal terms are absent at $T=0$, only $\Theta$-functions appear
on such lines, and these reproduce the familiar cuts.

Furthermore \cite{welcut}, the imaginary part of the self-energy
can be written as
\begin{equation}
   {\rm Im}\,\Pi(p)\;=\;p\cdot u\,\Gamma_r(p\cdot u)\;=\;
   p\cdot u(\Gamma-\Gamma_p),
\label{htl.relax}\end{equation}
where $\Gamma_r$ is the relaxation rate, $\Gamma_p$ is the production rate
of the particle in the heat bath and $\Gamma$ is the computed decay rate.

Now, for the decay rate, we require that the particles in the loop are
observed in the final state. We might guess that the computation requires
the replacement
\begin{equation}
   iG^\pm(p)\;=\;2\pi\delta(p^2-m^2)\Theta(\pm p_0)\left[1+B(p\cdot u)\right].
\label{htl.decay}\end{equation}
The overall factor of $\Theta$ allows us to talk of cut lines as at $T=0$.
Similarly, for the production rate we might guess that the replacement,
\begin{equation}
   iG^\pm(p)\;=\;2\pi\delta(p^2-m^2)\Theta(\mp p_0)\left[1+B(p\cdot u)\right],
\label{htl.prod}\end{equation}
is called for. Then do we miss cross terms which involve a $\theta(p_0)
\theta(-p_0')$? A simple check reveals that if the two-point function involves
a real on-shell particle, then such cross terms vanish due to kinematical
reasons. This is the content of eq.\ (\ref{htl.relax}), and the
justification for the prescription in eq.\ (\ref{htl.decay}), used in
Section \ref{temp}.

In our computation of the decay width, $\Gamma$,
we have neglected hard thermal loops \cite{htl}, corresponding to screening
of electric charges. Knowing that they are
important for computations of the photon self-energy at $T>0$, for 
$k\cdot u\sim eT$, is their neglect justified? The answer is: yes, because
these have large contributions to $\Gamma_r$, but their effect on $\Gamma_p$
and $\Gamma$ separately are small. We give the outline of a demonstration
below.

The hard thermal loop contributions to the process we are interested in
can be represented as
\begin{equation}
   \Gamma^{(ab)}\;=\;
   \begin{picture}(100,30)(-80,15)
      \DashLine(-80,20)(-50,20){3} \DashLine(80,20)(50,20){3}
      \GOval(-40,20)(20,10)(0){0.5} \GOval(40,20)(20,10)(0){0.5}
      \Photon(-34,35)(34,35){5}{3.5} \Photon(-34,5)(34,5){5}{3.5}
      \Text(-40,20)[c]{$a$} \Text(40,20)[c]{$b$}
      \Text(-32,42)[lb]{$\mu$} \Text(0,42)[b]{$k_1$} \Text(32,42)[rb]{$\sigma$}
      \Text(-27,17)[lt]{$\nu$} \Text(0,17)[t]{$k_2$} \Text(27,17)[rt]{$\rho$}
   \end{picture}
\label{htl.htl}\end{equation}
The blobs represent all vertex corrections to the triangle diagram and
self energy corrections to the Fermion and photon lines. The result is,
retaining only the terms of interest,
\begin{equation}\begin{array}{rl}
   \Gamma^{(ab)}\;=&\;\int d^4k_1 d^4k_2
           M^{(a)}_{\mu\nu} M^{(b)}_{\sigma\rho}\,
             {\rm Im}\Pi_{\mu\sigma}(k_1)\,{\rm Im}\Pi_{\nu\rho}(k_2)\\
       \;\sim&\;\int d^4k_1 d^4k_2 \left[1+B(k_1\cdot u)\right]
              \left[1+B(k_2\cdot u)\right]\\
         &\quad\times
            \biggl[ M^{(a)}_{ll} M^{(b)}_{ll} \rho_l(k_1)\rho_l(k_2)
                 + M^{(a)}_{lt} M^{(b)}_{lt} \rho_l(k_1)\rho_t(k_2)\\
         &\qquad + M^{(a)}_{tl} M^{(b)}_{tl} \rho_t(k_1)\rho_l(k_2)
                 + M^{(a)}_{tt} M^{(b)}_{tt} \rho_t(k_1)\rho_t(k_2)\biggr],
\end{array}\label{htl.cut}\end{equation}
where $\rho_l$ and $\rho_t$ are the spectral densities of the longitudinal
and transverse components of the photon. The quantities $M^{(a,b)}$ are the
contributions from the blobs; the subscripts denote appropriate components
of the tensors. The analysis of \cite{rob} can be
adapted to show that the term quadratic in $\rho_t$ is the dominant
infra-red term and gives the same contribution as that obtained with tree
level propagators for the photon. Hence, hard thermal loop resummation
does not change our results for $\Gamma$. Similiar arguments hold for the
production rate $\Gamma_p$. However, the leading terms, quadratic in $\rho_t$,
cancel between the production and decay rates and hence the sub-leading terms
become important for a reliable computation of the relaxation rate $\Gamma_r$.
For this reason, hard thermal loop resummation becomes important for
the computation of the radius and other properties of the axio-sphere.


\bigskip
\begin{thebibliography}{99}
\bibitem{axion}
   R.\ D.\ Peccei and H.\ R.\ Quinn, {\sl Phys.\ Rev.\ Lett.\/}, 38 (1977) 1440
   and {\sl Phys.\ Rev.\/}, D 16 (1977) 1791;\\
   S.\ Weinberg, {\sl Phys.\ Rev.\ Lett.\/}, 40 (1978) 223;\\
   F.\ Wilczek, {\sl Phys.\ Rev.\ Lett.\/}, 40 (1978) 279;\\
   J.\ E.\ Kim, {\sl Phys.\ Rev.\ Lett.\/}, 43 (1979) 103;\\
   M.\ Zhitnitskii, {\sl Sov.\ J.\ Nucl.\ Phys.\/}, 31 (1980) 260;\\
   M.\ A.\ Shifman, A.\ I.\ Vainshtein, V.\ I.\ Zakharov,
             {\sl Nucl.\ Phys.\/}, B 166 (1980) 493;\\
   M.\ Dine, W.\ Fischler and M.\ Srednicki, {\sl Phys.\ Lett.\/},
             B 104 (1981) 199.
\bibitem{cosmo}
   J.\ E.\ Kim, {\sl Phys.\ Rep.\/}, 150 (1987) 1;\\
   M.\ S.\ Turner, {\sl Phys.\ Rep.\/}, 197 (1990) 67;\\
   G.\ G.\ Raffelt, {\sl Phys.\ Rep.\/}, 198 (1990) 1.
\bibitem{weldon}
   H.\ A.\ Weldon, {\sl Phys.\ Rev.\/}, D 26 (1982) 1394.
\bibitem{htl}
   E.\ Braaten and R.\ Pisarski, {\sl Nucl.\ Phys.\/}, B 337 (1990) 569.
\bibitem{prev}
   C.\ Contreras and M.\ Loewe, {\sl Z.\ Phys.\/},  C 40 (1988) 253;\\
   Bi Pin-Zhen and J.\ Rafelski, {\sl Mod.\ Phys.\ Lett.\/},
     A 7 (1992) 2493;\\
   A.\ Gomez Nicola and R.\ F.\ Alvarez-Estrada, {\sl Z.\ Phys.\/},
     C 60 (1993) 711.
\bibitem{textbk}
   See for example, C.\ Itzykson and J.\ Zuber, {\sl Quantum Field
   Theory\/}, McGraw Hill Book Company, New York, 1985.
\bibitem{ftft}
   N.\ P.\ Landsman and Ch.\ G.\ van Weert, {\sl Phys.\ Rep.\/},
     145 (1987) 141.
\bibitem{dj}
   L.\ Dolan and R.\ Jackiw, {\sl Phys.\ Rev.\/}, D 9 (1974) 3312.
\bibitem{axion-soft}
   P.\ Sikivie, {\sl Phys.\ Rev.\ Lett.\/}, 48 (1982) 1156;\\
   M.\ S.\ Turner, {\sl Phys.\ Rev.\/}, D 33 (1986) 889;\\
   R.\ Davis, {\sl Phys.\ Lett.\/}, B 180 (1986) 225;\\
   D.\ Harari and P.\ Sikivie, {\sl Phys.\ Lett.\/}, B 195 (1987) 361;\\
   A.\ Dabholkar and J.\ M.\ Quashnock, {\sl Nucl.\ Phys.\/}, B 333 (1990) 815.
\bibitem{anjan}
   A.\ S.\ Joshipura, {\sl Int.\ J.\ Mod.\ Phys.\/}, A 7 (1992) 2021.
\bibitem{mahapatra}
   E.\ Kh.\ Akhmedov, Z.\ G.\ Berezhiani, R.\ N.\ Mohapatra and
   G.\ Senjanovi\'c, {\sl Phys.\ Lett.\/}, B 299 (1993) 90.
\bibitem{ellis}
   J.\ Ellis and K.\ A.\ Olive, {\sl Nucl.\ Phys.\/}, B 233 (1983) 252.
\bibitem{turner}
   M.\ S.\ Turner, {\sl Phys.\ Rev.\ Lett.\/}, 60 (1988) 1797;\\
   R.\ Mayle {\sl et al.\/}, {\sl Phys.\ Lett.\/}, B 203 (1988) 188;\\
   A.\ Burrows, M.\ S.\ Turner and R.\ P.\ Brinkmann, {\sl Phys.\ Rev.\/},
   D 39 (1989) 1020;\\
   R.\ Mayle {\sl et al.\/}, {\sl Phys.\ Lett.\/}, B 219 (1989) 515.
\bibitem{leut}
   P.\ Gerber and H.\ Leutwyler, {\sl Nucl.\ Phys.\/}, B 321 (1989) 387.
\bibitem{cutting}
   R.\ L.\ Kobes and G.\ W.\ Semenoff, {\sl Nucl.\ Phys.\/}, B 260 (1985) 714;\\
   R.\ L.\ Kobes and G.\ W.\ Semenoff, {\sl Nucl.\ Phys.\/}, B 272 (1986) 329;\\
   F.\ Gelis, preprint hep-ph/9701410.
\bibitem{welcut}
   H.\ A.\ Weldon, {\sl Phys.\ Rev.\/}, D 28 (1983) 2007.
\bibitem{rob}
   R.\ Pisarski, {\sl Physica\/}, A 158 (1989) 146.
\end{thebibliography}
\end{document}